\documentclass[a4 paper, 12pt]{article}
\usepackage[lmargin=0.8in,rmargin=0.8in,tmargin=0.8in,headsep=.5in]{geometry}

\usepackage{amsmath,amsfonts,amssymb,amsthm}

\numberwithin{equation}{section}

\newcommand{\edf}{\ {\mathop{=}\limits^{\rm def}}\ }

\begin{document}
\vspace*{0.5cm}
{\noindent\huge\bf  Motion of Spinning  Density Tensors in a Clifford Space}\\[0.15cm]
\begin{center}
{\author{}{Magd E. Kahil{\footnote{ Faculty of Engineering, Modern Sciences and Arts University, Giza, Egypt  \\
E.mail: mkahil@msa.edu.eg}}} and { Samah A. Ammar{\footnote{Women's colleague for Arts,Science and Education,Ain Shams University, Cairo, Egypt \\ E.mail: Samah.Ammar@women.asu.edu.eg}}} \\ {} } 
\end{center}

\begin{center}
  {\rule{0.9\textwidth}{0.5pt}}\\
\end{center}
{\bf\large Abstract}
 A Clifford Space is counted to be a tempting approach to unify both micro-physics and macro-physics simultaneously. Such a tendency may be found in the realm of replacing vectors with poly-vectors.  Accordingly, the problem of motion becomes essential to express the motion of extended particles rather than test particles. These equations are performed by using an equivalent Bazanski Lagrangian in a Clifford space.   From this perspective, a generalized type an equation for spinning density tensors and spinning density deviation tensors are obtained. Spinning deviation tensors in a Clifford space may give a better performance to examine the problem of stability for spinning density tensors as expressed in terms of vectors defined in such a class of Riemannian geometry.



\section{Introduction}
Geometrization of physics is counted to be one of the building blocks to perceive the philosophy of general relativity. This concept may lead to describing physics within geometry, this may give rise to continuously revisiting the notation of the type of physics as well as seeking earnestly to explore different types of geometries capable of expressing physics. \\
Accordingly, such a challenging issue becomes vital to examine, which is the geometrization of microphysics together with microphysics. This type of inquiry has focused on searching for an eligible type of geometry that can geometrize both branches of physics simultaneously. \\

Due to an insightful vision of microphysics, the question of the behavior of matter within a microscopic level has introduced the situation of gauge invariance with its strong, weak, and electromagnetic interactions that can be fundamentally invariant under local symmetry, while gravity is not combined with them.  Nevertheless, such a remedy to describe general relativity a gauge theory using inserting tetrad field and spin connection is regarded as gauge potential which can be found throughout its corresponding covariant derivative. Using derivatives one can obtain invariance under the general coordinate transformation (GCT) as well as Local Lorentz Transformation (LLT) by Collins et al \cite{Collins1989}. 
 However, this technique is unable to detect the intrinsic spin of the particle, such an approach has to amend   Riemannian geometry with non-Riemannian geometry, taking as a preliminary stage to the Riemann-Cartan geometry which embodies in its classes the well-known Einstein-Cartan space. The advantage of Einstein Cartan geometry is due to the existence of torsion tensor as being capable of describing intrinsic spin subject to Poincar\'{e} gauge theory \cite{Hehl1976},    \cite{Lasenby1998}.

Fortunately, it has been figured out the Clifford Space may work for unifying both micro and maco-physics simultaneously using Geometric Algebras while employing the concept of poly vectors rather than vectors. The following section briefly accounts for Clifford's space and its underlying geometry \cite{Castro2005}. \\ From this perspective, equations of motion for test particles will be replaced by extended ones as expressed in section (3) using a suggested method of obtaining both poly-vector and poly-vector deviation \cite{Kahil2020}.

Accordingly, the paper is organized as follows. In Section $2$, we display the fundamentals of Clifford space \cite{Castro2013}, Section 3 extends the Bazanski approach\cite {Bazanski1989, Kahil2006} for poly-vectors \cite{Kahil2020}

Consequently, Section 4 displays  a new approach of spinning fluid and the spinning density poly-tensor is presented as an extension to its analog in Riemaian geometry  \cite{Kahil2023}.  The aim of dealing with the behavior of spinning fluid and spinning density is a reliable issue to has been etended to examine the spinning fluid of a variable mass as shown in Section (6)   which is extended to introduce a generalized form of spin density for poly-vectors in which it is capable to describe the motion of spinning density matter such as plasma in   express how  near by the AGN  as well as in early  cosmology as illustrated in Section (7).\\
 Section (8), performs due to the importance of spin density deviation poly-tensor and its reliability, the stability conditions for a spinning density orbiting a compact object \cite{Kahil2015}.\\ Thus,  the stability problem of poly-tensor for  extended objects which is related to a deviation poly-vector, whether the stability is concerned with a test particle,  a spinning object or even a spinning density poly-tensor may be obtained. 

\section{ C-space: An Overview }

Following the geometrization scheme of physics, within the framework of geometric algebras rather than the well-known tensor formalism. From this perspective, an alternative elegant formalism has been performed by means of a Clifford space. The advantage of this space is its ability to explore new hidden physics or show an insightful vision for revisiting the old notations of physics. Such a trend is crystal clear using Clifford Algebra which leads to expressing many physical quantities in a compact form. It is worth mentioning that, physical objects are considered as matter in space-time, they can be informed of membranes (brane) of various dimensions (p-branes) that appear as a center of mass described by a point particle and extended objects in terms of closed strings, ...etc.  \cite{Pavsic2005}  \\
 Due to the richness of Clifford algebra is scalars, vectors, bi-vectors, and r-vectors are expressed in one form called Clifford aggregate or Poly-vector. Accordingly, it can be figured out that the Dirac matrix plays a fundamental role in combining, scalars, vectors, and tensors in one form called a Clifford aggregate. From this perspective, the coordinates o a poly-vector$ {X^{M}}$,  as follows \cite{Castro2005}  
 
\begin{equation}
X^{M} \equiv X^{\mu_{1}\mu_{2}\mu_{3}....}
\end{equation}
Accordingly, it can be shown that poly-vector coordinates of  C-space are parameterized not only by 1-vector coordinates $x^{\mu}$ but also by the 2-vector coordinates $x^{\mu \nu}$, 3-vector coordinates $x^{\mu \nu \rho}$ ...etc called holographic coordinates, such that
\begin{equation}
X^{M} =s {\bf{1}}+ x^{\mu}\gamma_{\mu} = x^{\mu \nu} \gamma_{\mu} \wedge \gamma_{\nu} +  x^{\mu \nu \rho} \gamma_{\mu} \wedge \gamma_{\nu} \wedge \gamma_{\rho} + x^{\mu \nu \rho \sigma} \gamma_{\mu} \wedge \gamma_{\nu} \wedge \gamma_{\rho} \wedge \gamma_{\sigma}
\end{equation}

where the component $S$ is the Clifford scalar components of a poly-vector valued coordinates. \\

Thus, in C-space proper time interval may be described as in Minkowski space \cite{Castro2013} \\
 
 \begin{equation}
 (d S)^{2} = (ds)^{2} + dx_{\mu}dx^{\mu}  + dx_{\mu \nu}dx^{\mu \nu}+....
 \end{equation}

Yet, it is vital to be noted that in Clifford Space, there is a striking virtue, unlike the conventional string theory  which expressed within 26 dimensions in which gauge fields are described as compact  dimensions ; while in C-space there is only 16 non-compact dimensions. Consequently, \cite{Pavsic2005}    generalized the conventional 4-dimensional space-time into 16 dimensional \`{a} la Klauza-Klein theory. \cite{Castro2005}. \\
 \\ (i) There is no need for extra-dimensions of space-time. Extra-dimensions in C-space.  \\
  (ii) There is no need to compactify the extra-dimensions. The extra-dimensions of C-space, namely $\omega, x^{\mu \nu}, x^{\mu \nu \rho}$,and $ x^{\mu \nu \rho \sigma}$ sample the extended objects, therefore they are physical dimensions. \\  (iii)The number of components $G_{\mu \tilde{M}}$ , $\mu=0,1,2,3$ and $\tilde{M} \neq \mu $ is 12, which is the same as the number of gauge fields in the standard models.  \\
Thus, the line element of poly-vectors in C-space become
\begin{equation}
|dX|^{2} = dS^2 =G_{MN}dX^{M}dX^{N}  \\
 \end{equation}
i.e.
$$
dS^2 = d s^2 + L^{-2}dx_{\mu}dx^{\mu}+ L^{-4}dx_{\mu \nu}x^{\mu \nu} +  L^{-6}dx_{\mu \nu \rho}x^{\mu \nu \rho}+....
$$
where $G_{MN} = {E}^{\dag}_{M}E_{N}$ is c-Space metric, $L$ is the Planck length \cite{Castro2005}.
equation (2.43)can be expressed as
\begin{equation}
 dS^2 = c^{2} dt^{2}( 1 + \frac{L^{-2}}{c^2} \frac{dx_{\mu}}{dt} \frac{dx^{\mu}}{dt}+ \frac{L^{-4}}{c^2}\frac{dx_{\mu \nu}}{dt}\frac{dx^{\mu \nu}}{dt} +  \frac{L^{-6}}{c^2}\frac{dx_{\mu \nu \rho}}{dt}\frac{dx^{\mu \nu \rho}}{dt}+....)
\end{equation}
with taking into account that  the parameter $s$ is  described only in the ordinary space-time i.e. $(d s)^{2} = g_{\mu \nu} dx^{\mu}dx^{\nu}$ .  

 Meanwhile, in C-space, Castro and Pavsic (2005)\cite{Castro2005}, showed that,  there are some principles must  be revisited especially the speed of light which is no longer the upper limit to be reached,  but another combination  of  constants made of Planck's length in the following way \cite{Castro2005}: \\
(i). Maximum 1-vector speed $\frac{d x^{\mu}}{d s} = 3 \times 10^8 m{s}^{-1}  $ ,\\
(ii) Maximum 2-vector speed $\frac{d x^{\mu \nu}}{d s} = 1.6 \times 10^{-35}m^{2}{s}^{-1}  $ ,\\
(iii) Maximum 3-vector speed $\frac{d x^{\mu \nu \rho}}{d s} = 7.7 \times 10^{-62} m^{3}{s}^{-1}  $ ,\\
(iv) Maximum 4-vector speed $\frac{d x^{\mu \nu \rho}}{d s} = 7.7 \times 10^{-96} m^{4}{s}^{-1}   $ ....etc  \\

Also, it is worth mentioning that using C-space a particle as observed from 4-dimensional space, that can be speed of particles may be more than the conventional speed of light due to the 8involvement of holographic coordinates $x^{\mu \nu}$, $x^{\mu \nu \rho} $...etc ; but does not exceed the modified speed of light due to C-space. This can lead us to consider there will be a specific upper limit to objects due to p-brans as $p=0,1,2,3,...$. 
The Line element of poly-vectors in C-space is defined as follows \cite{Castro2005}
\begin{equation}
dS^{2} = G_{MN} dX^{M} \times dX^{N}
\end{equation}
Such that the matrix $G_{MN}$ is defined as
\begin{equation}
 G_{MN} = \left[ \begin{array}{cc}
g_{\mu \nu} &  G_{\mu \bar{N}} \\
G_{ \bar{M}\nu} & G_{\bar{M} \bar{N}} \\
 \end{array} \right]
\end{equation}
Those degrees of freedom are in principle not hidden by which we sample the extended objects, therefore we do not need to compactify such internal space. \\
The metric of C-space $G_{MN}$ is subdivided into $G_{\mu \nu}=g_{\mu \nu}$ which relates to gravity, while gauge fields $G_{\mu \bar{M}}$, where $\mu \neq \bar{M}$ assumes 12 possible values, excluding the four values of $\nu =0,1,2,3$, and other 12 gauge fields ,to be defined as follows:  1 photon,3 weak gauge bosons and 8 gluons described  $A_{\mu}$,  $A^{a}_{\mu}, a=1,2,3$ and by $A^{c}_{\mu} c=1,1,3,...8.$ respectively.\\
It can be found that the number of mixed components in $G_{\mu \bar{M}} =(G_{\mu {\bf{1}}}$, $ G_{\mu [\alpha \beta]}$, $ G_{\mu [\alpha \beta \gamma]}$, $G_{\mu [\alpha \beta \gamma \delta]})$ of Clifford metric coincides with the number of gauge fields in the standard model. For a fixed ${\mu}$, there are 12 mixed components of $G_{\mu \bar{M}}$ and 12 gauge fields $A^{a}_{\mu}, W^{c}_{\mu} , A^{c}_{\mu}$. 
 This coincidence is fascinating and it may indicate that the known interactions are incorporated in curved Clifford space.\\ The number of mixed metric $G_{\mu \bar{M}}$ is 12 the same number as the number of gauge fields in the standard model.
,
Nevertheless, there are also interactions due to components $G_{\bar{M}\bar{N}}$ but do not have the properties of Yang-Mills, taking into consideration the idea presented by Pavsic (2006) who has considered it as a metric of an internal space \cite{Kahil2020} and references theirin.
From tis perspective, the Clifford version of bi-gravity theory may be constructed.  Such a study will be performed in our future work.   \\
\subsection{ C-space: Underlying Geometry}
{{ A point in C-Space is defined  as a set of holographic coordinates $(s,x^{\mu}, x^{\mu \nu},...)$ forming the coordinates of a poly-vector . Each one  is expressed within bases $\{\gamma_{A}\} =\{1, \gamma_{a_{1}},\gamma_{a_{1}a_{2}},\gamma_{a_{1}a_{2}a_{3}}....  \} $} , $a_{1}<a_{2}<a_{3}<a_{4}<a_{5}<...$, $r=1,2,3...., $ where $\gamma_{a_{1}a_{2}a_{3}...} = a_{1}\wedge a_{2}\wedge a_{3}....$ . \\ It is well known that the local basis $\gamma_{\mu}$, is related to the tetrad field $e^{a}_{\mu}$ such that $$\gamma_{\mu} = e^{a}_{\mu}\gamma_{a} $$.\\

   In a geometric product a scalar and a bi-vector occur in one sum. Clifford algebra is a superposition, called a Clifford aggregate or poly-vector.
   \begin{equation}
   A= a + \frac{1}{1!}a^{\mu}\gamma_{\mu} + \frac{1}{2!}a^{\mu \nu}\gamma_{\mu} \wedge  \gamma_{\nu}+..........\frac{1}{n!}a^{\mu_{1}.....\mu_{n}} \gamma_{\mu_{1}} \wedge ......\gamma_{\mu_{n}}
   \end{equation}
The differential of C-space is defined as follows  \cite{Castro2013}
\begin{equation}
d A= \frac{\partial A}{\partial X^{B}}dX^{B}
\end{equation}
i.e
\begin{equation}
d A= \frac{\partial A}{\partial s} +  \frac{\partial A}{\partial x^{\mu}}dx^{\mu}+  \frac{\partial A}{\partial x^{\mu \nu}}dx^{\mu \nu}+......
\end{equation}
in other words, if one takes $A =\gamma_{\mu}$ , then
\begin{equation}
d \gamma_{\mu}= \frac{\partial \gamma}{\partial s} +  \frac{\partial \gamma_{\mu}}{\partial x^{\mu}}dx^{\mu}+  \frac{\partial \gamma_{\mu}}{\partial x^{\mu \nu}}dx^{\mu \nu}+......
\end{equation}
which becomes

\begin{equation}
d \gamma_{\mu}= \frac{\partial \gamma}{\partial s} +  \Gamma^{\alpha}_{\mu \nu }dx^{\nu}+  \Gamma^{\alpha}_{\mu [\nu , \rho] }dx^{\nu \rho}+......
\end{equation}
which can be reduced to 
\begin{equation}
d \gamma_{\mu}= \frac{\partial \gamma}{\partial s} +  \Gamma^{\alpha}_{\mu \nu }dx^{\nu}+  \frac{1}{2}R^{\alpha}_{\beta \nu \rho}dx^{\nu \rho}+......
\end{equation}
where $R^{\alpha}_{\beta \nu \rho}$ is the curvature of space-time. \\
Also, it can be found that for an arbitrary poly-vector $A^{M}$ 
\begin{equation}
\frac{D A^{M}}{D x^{\mu \nu}} = [D_{\mu}, D_{\nu}] A^{M}
\end{equation}
where $\frac{D}{D x^{\mu \nu}}$ is the covariant derivative with respect to a plane$x^{\mu \nu}$,  such that

\begin{equation}
\frac{D s}{D x^{\mu \nu}} = [D_{\mu}, D_{\nu}] s= K^{\rho}_{\mu \nu} \partial_{\rho}s,
\end{equation}
where $K^{\rho}_{\mu \nu}$ is the torsion tensor.
\begin{equation}
\frac{D a^{\alpha}}{D x^{\mu \nu}} = [D_{\mu}, D_{\nu}] a^{\alpha}=  R^{\alpha}_{\rho \mu \nu} a^{\rho}+ K^{\rho}_{\mu \nu} D_{\rho}a^{\alpha},
\end{equation}
Yet, this type of torsion (2.12) can related to the notation of torsion as mentioned by Hammond as commutator the potential associated by a prescribed scalar field $\phi$ [25]
\begin{equation}
 K^{\alpha}_{\mu \nu} = \frac{1}{2}(\delta^{\alpha}_{\mu} \phi_{,\nu} -\delta^{\alpha}_{\nu} \phi_{,\mu} )
\end{equation}
 Thus, we can figure out that the torsion as defined in C-space (Riemannian Type) through the parameters $s$ may act like an independent scalar field defined in the usual context of Riemannian-Cartan geometry.

From examining equations (2.11) and (2.12), one can find the existence of a torsion tensor in the even presence of a symmetric affine connection apart from its conventional notation definition of being the antisymmetric part of an affine connection as in the context of non-Riemannian geometries \cite{Pavsic2005}.  Accordingly, owing to C-space one may realize the dispute between the reliability of torsion propagating or non-propagating this can be resolved by means of of we describe torsion as a result of covariant differentiation of areas of holographic coordinates and non-propagating as being defined as anti-symmetric parts of an affine connection of poly-vectors or vectors, if one utilizes in the internal or external coordinate its corresponding non-symmetric affine connection. 

Consequently, we can regard that the geometry described within the coordinates of poly-vector described not only Riemannian but also a non-Riemannian, i.e. the composition of a Riemannian affine connection for the poly-vector is not necessarily Riemannian as well, This may through some light to find out that a Riemannian poly-vector affine connection and curvature (external coordinate  capital Latin letters) may be described by non Riemannian quantities as describing its holographic coordinates (internal coordinate (Greek letters)) \cite{Kahil2020}. \\

{\underline{(Curved Clifford Space)}}
  Yet, in C-space it is convenient to distinguish two frame fields \cite{Castro2005}: 
   \\ (i) Coordinate frame field , whose bases elements $\Gamma_{M}, M=1,2,3....2^{n}$ depend on the position of $X$ in C-space such that its the relation of wedge can not be preserved globally . 
\begin{equation}
  \gamma_{M} \equiv  \gamma_{\mu_{1}....\mu_{r}},
  \end{equation}
This leads to the scalar product of them gives the metric tensor of the C-space \\
      \begin{equation} 
 ^{\ddag}{\gamma}_{M} . \gamma_{N} =G_{MN},
\end{equation}
Such that its associate covariant derivative is defined as 
\begin{equation}
  \partial_{M} \gamma_{N}  = \Gamma^{J}_{MN} \gamma_{J},
  \end{equation}
   \\ (ii) Orthonormal frame field, whose bases elements $\gamma_{(A)}$, $(A)=1,2,3,...2^{}$ depend on X, such that at every point there are wedge products determine  \\
   \begin{equation}
  \gamma_{(A)} \equiv  \gamma_{a_{1}....a_{r}} = \gamma_{a_{1}} \wedge \gamma_{a_{2}}\wedge \gamma_{a_{3}}......\wedge \gamma_{a_{r}},
  \end{equation}
  
  \begin{equation} 
 ^{\ddag}{\gamma}_{(A)} . \gamma_{(B)} =\eta_{AB} 
\end{equation}
where $\eta$  is the metric tensor in flat space.\\
  The derivative of a poly-vector is classified as follows \cite{Casrto2002}:
  \\({i){\underline{Scalar field}} \\
 It is behaving as an ordinary partial derivative. i.e
  \begin{equation}
  \partial_{M} \phi = \frac{\partial \phi}{\partial X^{M}}.
  \end{equation}
  (ii){\underline{Coordinate bases}} \\
 On generic poly-vector field ,becomes
  \begin{equation}
  \partial_{M} \gamma_{N}  =  \Gamma^{Q}_{MN} \gamma_{Q} 
  \end{equation}
 where $\Gamma^{Q}_{MN}$ is an affine connection, and the commutator of its derivative be
\begin{equation}
[\partial_{M}, \partial_{N}]\gamma_{J} = R^{K}_{.~JMN}\gamma_{K} 
\end{equation}
(iii) {\underline{Orthonormal bases (local frame field)}} \\
it turns to be  
\begin{equation}
  \partial_{M} \gamma_{(A)}  = - \Omega_{A~.~M}^{N}\gamma_{(B)} 
  \end{equation}
such that $\Omega_{A~.~M}^{N}$  acts as its appropriate spin connection. \\ 
Thus,  the commutator of its derivative for the coordinate frame field becomes
$\Gamma^{Q}_{MN}$ is an affine connection, and 
\begin{equation}
[\partial_{M}, \partial_{N}]\gamma_{A} = R^{B}_{.~AMN}\gamma_{B} .
\end{equation} 
Thus ,  one may define the covariant derivative for an aggregate poly-vector  $A$ one finds that:
\begin{equation}
  \partial_{M} A^{N}\gamma_{N}  = \partial_{M}A^{N}\gamma_{N} + A^{N} \partial_{M}A^{N} = (\partial_{M} A^{M}+ \Gamma^{N}_{MK} A^{K}) \gamma_{N} \equiv D_{M}A^{N}\gamma_{N}
  \end{equation}
\subsection{C-space and The Tetrad Field}
The relationship between the two bases are related by the tetrad field $E^{A}_{}$ associated with poly-vectors [24]
  \begin{equation}
  \gamma_{M} = E^{(A)}_{M} \gamma_{(A)},
  \end{equation}
  \begin{equation}
  \partial_{M} \phi = \frac{\partial \phi}{\partial X^{M}}.
  \end{equation}
  where $ E^{A}_{M} $ is the C-space vielbein,
Such that
\begin{equation}
 G_{MN} = E^{(A)}_{M}E^{(B)}_{N} \eta_{AB}
\end{equation}
such that $\eta_{AB}= \gamma_{(A)}\gamma_{(B)}$
whose covariant derivative

Such a description does not preclude the Non-Riemannian version of poly-vectors, which is a step to revisit the definition of the fundamental quantities that play an active role of describing such a type of geometry.

Accordingly, it is vital to note that the covariant derivative of C-space associated with Vielbein of poly-vector takes the following condition \cite{Kahil2020}
\begin{equation}
 \partial_{N} E^{C}_{M} - \Gamma^{P}_ {M N }E^{C}_{P} - E^{A}_{M}\Omega_{A~.~N}^{C} =0,  
\end{equation}
Consequently, the covariant derivative  shows that the tetrad field is invariant under general coordinate transformation through $\Gamma^{M}_{NQ}$  and local Lorentz invariance as expressed in terms of $\Omega_{A~.~N}^{C}$ represents spin connection in C-space in which it is invariant under the general coordinate transformation of poly-vectors the connection for the orthonormal frame field \cite{Castro2013}
  Similarly, its associated covariant derivative is 
  The components $D_{M} A^{N}$ are covariant derivative of tensor analysis

  The curvature of C-space is defined, as usually, by the commutator of derivatives acting on basis poly-vectors ]:
  \begin{equation}
  [D_{M}, D_{N}] \gamma_{J} = {R}^{K}_{MNJ}\gamma_{K}
  \end{equation}
  or
  \begin{equation}
  [D_{M}, D_{N}] \gamma_{(A)} = {R}^{(B)}_{MNJ}\gamma_{(B)}
  \end{equation}
where $D_{M}$ is the covariant derivivative.

Meanwhile, introducing the reciprocal basis poly-vectors $\gamma^{M}$ and $\gamma^{A}$ satisfying
  \begin{equation}
  (\gamma^{M})^{\ddag}* \gamma_{N} =\delta^{M}_{N} ,
  \end{equation}
  \begin{equation}
  (\gamma^{(A)})^{\ddag}* \gamma_{(B)} =\delta^{(A)}_{(B)} ,
  \end{equation}
  The components of  curvature in the corresponding basis
 \begin{equation}
 R^{K}_{MNJ} = \partial_{M} \Gamma^{K}_{NJ} - \partial_{N} \Gamma^{K}_{MJ} + \Gamma^{L}_{NJ}\Gamma^{K}_{ML}   -\Gamma^{L}_{MJ}\Gamma^{K}_{NL}
 \end{equation}
  or
  \begin{equation}
 {R}^{K}_{MN(A)} = \partial_{M} \Omega^{~K~}_{(A)~N} - \partial_{N} \Omega^{~K~}_{(A)~M} +  \Omega^{~(C)~}_{(A)~N}\Omega^{~(B)~}_{(C)~M}- \Omega^{~(C)~}_{(A)~M}\Omega^{~(B)~}_{(C)~N},
 \end{equation}
which is the analogous covariant derivative of the bases of vector $e^{c}_{\mu}$   in Riemannian geometry for manifolds of points-like \cite{Castro2005}.\\
Moreover, there is a counterpart version of non-vanishing curvature and torsion as defined in non-Riemannian geometry, in C-space leading to define torsion of poly-vectors to become
\begin{equation}
\Lambda^{J}_{MN} = \bar{\Gamma}^{J}_{MN} -\bar{\Gamma}^{J}_{NM}.
\end{equation}
Also, the contortion of poly-vectors $\Omega_{BCM}$ is given by 
\begin{equation}
\Omega_{BCM} = \frac{1}{2}E^{(A)}_{M}(\Delta_{[(A)(B)](C)} -\Delta_{[(B)(C)](A)} + \Delta_{[(C)(A)](B)} )
\end{equation}
where  $\Delta_{[(A)(B)](C)}$ is defined as the Ricci coefficient of rotation 
$$
\Delta_{[(A)(B)](C)} = E^{M}_{(A)}E^{N}_{(B)}(\partial_{M}E_{N(C)} - \partial_{M}E_{N(C)} ).
$$
Moreover, another covariant derivative associated with non-symmetric affine connection $\bar{\Gamma}^{M}_{NK} $  is defined as follows
\begin{equation} 
X^{M}_{| N} = \partial_{N} X^{M} + \bar{\Gamma}^{M}_{NS} X^{S},
\end{equation}
 such that 
\begin{equation}
 \bar{\Gamma}^{M}_{NS} X^{S} = {\Gamma}^{M}_{NS} X^{S} + \Delta^{M}_{NS}
\end{equation}

\begin{equation}
     X^{M}_{| M S } - X^{M}_{| N S} =  \bar{R}^{M}_{QNS}X^{Q} +  \Lambda^{Q}_{NS} X^{M}_{| Q}.
\end{equation}
such that
\begin{equation}
 R^{K}_{MNJ} = \partial_{M} \bar{\Gamma}^{K}_{NJ} - \partial_{N} \bar{\Gamma}^{K}_{MJ} + \bar{\Gamma}^{L}_{NJ}\bar{\Gamma}^{K}_{ML}   -\bar{\Gamma}^{L}_{MJ}\bar{\Gamma}^{K}_{NL}
\end{equation}
Due to richness of these quantities this type work will be going to examine the behavior of extended objects subjects have sensitivity to these quantities in our future work, while we focus in our present study an approach to derive the equations of motion and their deviation paths for different extended objects spinning and charged for poly-vectors defined within the context of Riemannian-like C-Space  are displayed in the following sections . 

\section{The Bazanski Approach for Poly-vectors}
 Equations of geodesic and geodesic deviation equations Riemannian geometry are required to examine many problems of motion for different test particles in gravitational fields. This led many authors to derive them by various methods, one of the most applicable ones is the Bazanski approach \cite{Bazanski1989}  in which from one single Lagrangian one can obtain simultaneously equation of geodesic and geodesic deviations which has been applied in different theories of gravity , \cite{Kahil2006, Kahil2017, Kahil2018a, Kahil2018b, Kahil2020a} . Thus, by analogy, this technique in the case of Polyvector becomes \cite{Kahil2020},   
 \begin{equation}
 L = G_{MN} U^{M} \frac{D \Psi^{N}}{D S}
 \end{equation}
where, $G_{MN}$ is the metric tensor, $U^{M}$, is a unit tangent poly-vector of the path whose parameter is $S$, and $\Psi^{\nu}$ is the deviation poly-vector associated to the path $(S)$, $ \frac{D}{DS}$ is the covariant derivative with respect to parameter $S$.\\
Applying the Euler Lagrange equation, by taking the variation to the deviation poly-vector $\Psi^{C}$  
\begin{equation}
\frac{d}{dS} \frac{\partial L}{\partial \dot{\Psi}^{C}}- \frac{\partial L}{\partial {\Psi}^{C}} =0 ,
 \end{equation}
to obtain the geodesic equation
\begin{equation}
\frac{D U^{C}}{D S} = 0,
 \end{equation}
and taking the variation with respect the the unit poly-vector $U^{C}$,
\begin{equation}
\frac{d}{dS} \frac{\partial L}{\partial{U}^{C}}- \frac{\partial L}{\partial {x}^{C}} =0,
 \end{equation}
to obtain the geodesic deviation equation,
\begin{equation}
\frac{D^2 {\Psi}^{E}}{D S^2} = R^{E}_{A B C} U^{A}U^{B}\Psi^{C},
 \end{equation}
where $ R^{A}_{ BC D }$ is Riemann-Christoffel tensor described by poly-vectors.
\section{Spinning Motion in a Clifford Space}
Spinning motion is regarded as one of the actual features of the characteristic behavior of objects in nature, which led many authors to focus on the cause of the spinning process. Would be eligible to include its internal properties or discard them as a step of simplification? From this perspective, it is vital, to begin with equations of motion of spinning Mathisson-Papapetrou \cite{Papapetrou1951}
\begin{equation}\label{uniqIDw001}
\frac{D P^{A}}{DS} = \frac{1}{2} R^{A}_{.~BCD}S^{CD} U^{B}
\end{equation}
   where $P^{A}$ is the momentum of the particle
    $$P^{A}= (mU^{A}+ U_{M} \frac{D S^{AM} }{DS}),$$  $R^{A}_{.~BCD}$ is the Riemannian curvature, $U^{A}= \frac{dX^A}{dS}$ is the unit tangent vector, $S$ is a parameter varying along the curve and  $S^{CD}$ is the spin tensor. For a spinning object with precession (Gyroscopic motion) can be described using the following equation
\begin{equation}\label{uniqIDw002}
\frac{D S^{MN}}{DS} = P^{M}U^{N} -P^{N}U^{M}
\end{equation}
If $P^{A}= mU^{A}$ then equation \eqref{uniqIDw001} and \eqref{uniqIDw002} become, the Papapetrou equation for short!
\begin{equation}\label{uniqIDw003}
\frac{D U^{A}}{DS} = \frac{1}{2m} R^{M}_{BCD}S^{CD} U^{B},
\end{equation}
and
\begin{equation}\label{uniqIDw004}
\frac{D S^{MN}}{DS} = 0,
\end{equation}
provided that \cite{Kahil2018b}
\begin{equation}\label{uniqIDw005}
S^{\mu \nu} = \bar{S} ( U^{M} \Psi^{N} - U^{N} \Psi^{M}  )
\end{equation}
where, $\bar{S}$ spin magnitude, $U^{A}$ is four vector velocity and $\Psi^{A}$ is a geodesic deviation vector.
Equation \eqref{uniqIDw003} can be obtained from geodesic equations \cite{Bazanski1989}
\begin{equation}\label{uniqIDw006}
\frac{{DU}^{A}}{D W} =0
\end{equation} and the geodesic deviation equations
\begin{equation}\label{uniqIDw007} \frac{D^2 \Psi^{\alpha}}{D \tau^2} = R^{\alpha}_{~\mu \nu \rho} U^{\mu} U^{\nu} \Psi^{\rho}.  \end{equation}
If we apply the following transformation of paths for different parameters \cite{Kahil2018b}
\begin{equation}\label{uniqIDw008}
\frac{ dX^{A}}{dS} = \frac{dX^{A}}{d W} + \beta \frac{D \Psi^{A}}{D W}
\end{equation}
where $\beta \edf \frac{\bar{s}}{m}$ and by operating covariant derivative with respect to the parameter $S$ on both sides, we get
\begin{equation}\label{uniqIDw009}
\frac{D U^{A}}{DS} = \frac{D U^{A}}{DW} + \beta \frac{D^{2} \Psi^{A}}{D W^{2}}
\end{equation}
Using geodesic equations \eqref{uniqIDw006}  and geodesic deviation equations \eqref{uniqIDw007} as well Equation\eqref{uniqIDw003} one can obtain Equation \eqref{uniqIDw001}, while if one consider Frenkel condition
$$ S_{MN}U^N{M} =0$$ to be covariant differentiated on both sides and after some manipulations, one can get Equation\eqref{uniqIDw004}.

 Thus, due to the extension of spin tensor from pole-dipole moments to multi-pole moments for extended objects, this may lead to examining its corresponding propagation equation \cite{Yasskin1980}. Such an equation may be obtained by introducing the spin tensor density$S^{\alpha \beta \gamma}$ as a third order skew-symmetric tensor is viable to describe extended objects, in Clifford spaces by defining Spin density poly-tensor$S^{ABC}$ as a third order skew-symmetric poly- tensor. These equations play a vital role in revisiting astrophysics and early cosmology to become a good candidate for describing a spinning fluid and also for describing the status of the accretion disc orbiting a compact gravitational field as in AGN \cite{Kleidis2000}. Also, it contributes to understanding the problem of motion quark-gluon heavy ion collisions in the early universe \cite{Cao2022}.

In our present work we are going to extended what has been  derived from equations of spinning density tensor and spin deviation density  in GR \cite{Kahil2023} into their counterparts' equations expressed in a Clifford space.
Moreover, due to the interrelation between the spinning density tensor and thermodynamics variables\cite{Chrohok2002}, we may obtain its corresponding relation in a Clifford space to become.
\begin{equation}\label{uniqIDw010}
Td\hat{S}= d E + p d(\frac{1}{\rho})-\frac{1}{2}\Omega_{\mu \nu}dS^{M}
\end{equation}
where $T$ is the temperature, $\hat{s}$ is the entropy, $ E$ is energy density , $\Omega_{MN}$ is the spin angular velocity and $S^{MN}$ spin density. Thus, in self consistent theories described the entropy becomes conserved i.e.
such that
\begin{equation}\label{uniqIDw011}
\frac{d \hat S}{ dS} =1.
\end{equation}
Thus, the first law of thermodynamics becomes
\begin{equation}\label{uniqIDw012}
\frac{d E}{dS} + p \frac{d{\rho}^{-1}}{dS} - \frac{1}{2}\Omega_{MN} \frac{ds^{MN}}{dS} =0.
\end{equation}
 Meanwhile, it is worth to clarify that thermodynamics transports coefficients such as viscosity which is connected with identifying the nature of the spin tensor \cite{Becattanini2019}. It is well known that spinning fluids are dominating properties of nature, may be found to describe the problem of motion of particles in an accretion disc as a Gyrodynamics fluid. This is the counterpart of the Papapetrou equation \cite{Mosheni2008}. Owing to spin density tensor, an interaction between spinning motion and thermodynamics variables may be found in equation \eqref{uniqIDw012} \cite{Ray1982}.

Thus, it is well known that $S^{ABC}$ is a third order tensor, viable to define extended objects.
Nevertheless, from a progenitor case, the spin density tensor is bounded to be a skew symmetric in the last two indices, it comes to arise for expressing a spinning fluid element as a confided case of an extended object. From this perspective, the notation of importing the expression  propagating equation for spin density \cite{Yasskin1980} the modified becomes crystal clear to express spin fluids \cite{Chrohok2002}. Yet, one may find out that the Weyssenhoff tensor \cite{Cao2022} is most eligible candidate to express a spin fluid element i.e.
\begin{equation}\label{uniqIDw013}
{S}^{RMN} = S^{MN} U^{R}.
\end{equation}
 Thus, differentiating both  sides of \eqref{uniqIDw013} by the covariant derivative to get
\begin{equation}\label{uniqIDw014}
\frac{D{S}^{RMN}}{DS} = \frac{D S^{MN}}{DS} U^{R} + \frac{D U^{R}}{DS} S^{MN}.
\end{equation}
Now, one uses the following Lagrangian \cite{Kahil2018a}
\begin{equation}\label{uniqIDw015}
L= g_{MN} U^{M} \frac{D \Psi^{N}}{DS} + S_{MN}\frac{D \Psi^{MN}}{DS}.
\end{equation}
By taking the variation with respect to the deviation vector $\Psi^{R}$
one gets \eqref{uniqIDw006},
$$\frac{D U^{R}}{D0} =0.$$
Also, taking the variation with respect to the spinning deviation tensor $\Psi^{\rho \delta}$ to get \eqref{uniqIDw004}
\begin{equation}\label{uniqIDw016}
\frac{D S^{RD}}{DS} =0.
\end{equation}
Consequently, substituting from \eqref{uniqIDw006} and \eqref{uniqIDw016} into \eqref{uniqIDw014}, one obtain
\begin{equation}\label{uniqIDw017}
\frac{D S^{RDS}}{DS} =0,
\end{equation}
which indicates equation of spin density tensor.
Accordingly, we may obtain equations analogously by means of its corresponding Bazanski Lagrangian stemmed from its original formalism \cite{Bazanski1989} and its modification in GR to become
\begin{equation}\label{uniqIDw018}
L=  S_{AMN}\frac{D \Psi^{AMN}}{DS},
\end{equation}
such that by taking the variation with respect to $\Psi^{RDL}$,
\begin{equation}\label{uniqIDw019}
\frac{D S^{RDL}}{DS} =0.
\end{equation}
If we apply the commutation relation in such that
\begin{equation}\label{uniqIDw020}
(S^{CMN}_{~~~;AB }- S^{CMN}_{~~~;BA }) U^{A}\Psi^{B} = S^{S[ MN}R^{C]}_{~SAB} U^{A}\Psi^{B},
\end{equation}
and
\begin{equation}\label{uniqIDw021}
S^{CMN}_{~~~; D } \Psi^{D} =  \Psi^{CMN}_{~~~; D} U^{D}.
\end{equation}
Then we obtain its corresponding spin density deviation tensor equation
\begin{equation}\label{uniqIDw022}
\frac{D^{2} \Psi^{SDL}}{Ds^{2}} =S^{F [DL }R^{S ]}_{~FAB} U^{A} \Psi^{B}.
\end{equation}

\subsection{Spinning Density Tensor and Spinning Density Deviation Tensor Equations: Papapetrou-Like Equations}

In this section, we are going to examine a massive density spin tensor able to describe an orbiting extended object for a compact object. Accordingly, the Weyssenhoff spin vector may be amended to be expressed as follows:

\begin{equation}\label{uniqIDw201}
\bar{S}^{CMN} = S^{MN} P^{C},	
\end{equation}
where $P^{C}$ is the momentum in which it is relating to $\bar{S}^{CMN}$ in the following sense
\begin{equation}\label{uniqIDw202}
\bar{S}^{CMN} = S^{MN} (m U^{C}+ U_{D}\frac{D S^{CD}}{DS} ),
\end{equation}
i.e.
\begin{equation}\label{uniqIDw203}
\bar{S}^{CMN} = S^{MN} (m U^{C}+ U_{D}(P^{C} U^{D}- P^{D} U^{C} ) ).
\end{equation}
 Differentiating both sides by covariant derivative for \eqref{uniqIDw201} to get
\begin{equation}\label{uniqIDw204}
\frac{D \bar{S}^{CMN}}{DS} = \frac{D S^{MN}}{DS} P^{C} + \frac{D P^{C}}{DS} S^{MN}.
\end{equation}
We suggest the equivalent Bazanski Lagrangian to be:
\begin{equation}\label{uniqIDw205}
 L=  \bar{S}_{CMN} \frac{D \bar{\Psi}^{CMN}}{{D}{S}} + {f}_{CMN}\bar{\Psi}^{CMN}.
 \end{equation}
Thus, by taking the variation with respect to its corresponding deviation tensor $\bar{\Psi}^{CMN}$, we get
\begin{equation}\label{uniqIDw206}
\frac{D \bar{S}^{CMN}}{DS}= f^{CMN},
\end{equation}
in which $$f^{CMN}=\frac{D S^{MN}}{DS} P^{C} + \frac{D P^{C}}{DS} S^{MN},$$
becomes
$$f^{CMN}=(P^{M} U^{N} - P^{N}U^{M}) P^{C} + \frac{1}{2} R^{C}_{~ ASB} S^{ASB}  S^{MN},$$
where  $$ f^{M}= \frac{1}{2} {R}^{M}_{~NCD} S^{NCD},$$ is regarded as a spin force,  and $$ M^{MN} =P^{M}U^{N}- P^{N}U^{M}.$$
Consequently, its corresponding spin density deviation tensor equation can be obtained by applying in a similar way the commutation relations as given in \eqref{uniqIDw021} and \eqref{uniqIDw022} the corresponding spinning density deviation equation:
\begin{equation}\label{uniqIDw207}
\frac{{D}^{2}\Psi^{CMN}}{DS^{2}}=  S^{D[MN }{\bar{R}}^{C]}_{~ D AB} U^{A} \Psi^{B}+ \bar{f}^{CMN}_{~~~{;} D} \Psi^{D}.
 \end{equation}

 \subsection{Spinning and Spinning Deviation Equations of C-space}
We suggest the equivalent Bazanski Lagrangian for deriving the equations for spinning and spinning poly-vectors to be 
\begin{equation}
 L= G_{AB} P^{A} \frac{D \Psi^{B}}{D\bar{S}} + S_{AB} \frac{D \Psi^{AB}}{D\bar{S}}+ F_{A}\Psi^{A}+ M_{AB}\Psi^{AB}.
 \end{equation}
 such that $$ P^{A}= m U^{A}+ U_{\beta} \frac{D S^{A B}}{DS}$$ 
where $P^{\mu}$ is the momentum poly-vector $ F^{\mu} = \frac{1}{2} R^{\mu}_{\nu \rho \delta} S^{\rho \delta} U^{\nu}$, $R^{\alpha}_{\beta \rho \sigma}$ is the Riemann curvature, $\frac{D}{D\bar{S}}$ is the covariant derivative with respect  to a parameter $S$,$S^{\alpha \beta}$ is the spin poly-tensor, and $ M^{\mu \nu} =P^{\mu}U^{\nu}- P^{\nu}U^{\mu}$
 such that $U^{\alpha}= \frac{d x^{\alpha}}{ds}$ is the unit tangent poly-vector to the geodesic one.
 In a similar way as performed in (4.26), by taking the variation with respect to $ \Psi^{\mu}$ and$\Psi^{\mu \nu}$ simultaneously one obtains 
 \begin{equation}
\frac{DP^{M}}{D\bar{S}}= F^{M},
 \end{equation}
 \begin{equation}
\frac{DS^{M N}}{D\bar{S}}= M^{M N} ,
 \end{equation}
  \\
 Using the following identity on both equations (4.31) and (4.32)
  \begin{equation}
  A^{D}_{; N H} - A^{D}_{; H N} = {R}^{D}_{ B N H} A^{B},
  \end{equation}
  such that $A^{D}$ is an arbitrary poly-vector.
 Multiplying both sides with arbitrary poly-vectors, $U^{H} \Psi^{B}$ as well as using the following condition 
 \begin{equation}
 U^{A}_{; H} \Psi^{H} =  \Psi^{A}_{; H } U^{H},
 \end{equation}
and $\Psi^{A}$ is its deviation poly-vector associated to the  unit poly-vector tangent $U^{A}$.
 Also in a similar way:

 \begin{equation}
 S^{AB}_{; H} \Psi^{H} =  \Psi^{AB}_{; H } U^{H},
\end{equation}

 one obtains the corresponding deviation equations which are inspired from the workings of Mohseni \cite{Mosheni2008}
  \begin{equation}
\frac{D^{2}\Psi^{A}}{DS^{2}}=  \bar{R}^{A}_{B H C}P^{B} U^{H} \Phi^{C}+ F^{A}_{; H} \Psi^{H},
 \end{equation}
and
 \begin{equation}
\frac{D^{2}\Psi^{A B}}{D\bar{S}^{2}}=  S^{[B D }{R}^{A ]}_{ D H C} U^{H} \Psi^{C}+ M^{AB}_{; H} \Psi^{H}.
 \end{equation}
\subsection{The Generalized Dixon Equation in C-space }
As a byproduct, we suggest the following Lagrangian which enables us to obtain the spinning charged poly-vector in a Clifford space.
\begin{equation}
 L= G_{AB} P^{A} \frac{D \Psi^{B}}{DS} + S_{AB} \frac{D \Psi^{AB}}{DS}+ \Psi^{A}U^{B}+ \tilde{M}_{AB} \Psi^{AB},
 \end{equation}
such that $$ \tilde{F_{A}}= \frac{1}{m}(e F_{AB} + \frac{1}{2}{R}_{ABCD}S^{CD})$$ and  $$\tilde{M}_{AB}= (P_{A}U^{B}-P_{B}U^{U} +  F_{AC}S^{C}_{.~B}-F_{AC}S_{A~.}^{.~C} )$$ .
 Taking the variation with respect to $ \Psi^{\mu}$ and$\Psi^{\mu \nu}$ simultaneously we obtain
 \begin{equation}
\frac{DP^{M}}{DS}= \frac{\epsilon}{M} F^{M}_{B}U^{B} + \frac{1}{2M} {R}^{M}_{NEQ} S^{EQ}U^{N},
 \end{equation}
 \begin{equation}
\frac{DS^{M N}}{DS}= P^{M}U^{N}-P^{N}U^{M} + F^{MC}S_{C}^{.~N} - F^{CN}S^{M}_{.~C} .
 \end{equation}
Thus, applying the same laws of commutation as illustrated in (4.39)and (4.40) we obtain their corresponding deviation equation
  \begin{equation}
\frac{D^{2}\Psi^{A}}{DS^{2}}=  {R}^{A}_{B H C}P^{B} U^{H} \Phi^{C}+ ( \frac{\epsilon}{M} F^{M}_{B}U^{B} + \frac{1}{2m}{R}^{M}_{NEQ} S^{EQ}U^{N})_{; H} \Psi^{H},
 \end{equation}
and
 \begin{equation}
\frac{D^{2}\Psi^{A B}}{DS^{2}}=  S^{[B D }{R}^{A ]}_{ D H C} U^{H} \Psi^{C}+ (P^{M}U^{N}-P^{N}U^{M} + F^{M C}S_{C}^{.~ N} - F^{CN}S^{M}_{.~ C})_{; H} \Psi^{H}.
 \end{equation}
\section{Spinning Density Tensor and Spinning Density  Deviation Tensor Equations: A Variable Mass }
If we consider a massive spin density tensor  whose mass is not constant but function of the parameter $(S)$ in which its corresponding Weyssenhoff tensor becomes as follows
\begin{equation}\label{uniqIDw301}
\hat{S}^{CMN} = m(S)U^{C} S^{MN}.
\end{equation}
Differentiating both sides we obtain:
\begin{equation}\label{uniqIDw302}
\frac{D\hat{S}^{CMN}}{DS} = \frac{D(m(S)U^{C})}{DS} S^{MN} + m(S)U^{C}\frac{DS^{MN}}{DS}.
\end{equation}
Thus,  we can suggest a Lagrangian obtained for spinning variable mass object
\begin{equation}\label{uniqIDw303}
L= m(S)G_{MN}U^{\mu} \frac{D \Psi^{N}}{DS} + ( m(S)_{D} + \frac{1}{2} R_{D} S^{ABC})\Psi^{D} + S_{AB} \frac{D \Psi^{ABC}}{DS}+ m(S)_{D}\Psi^{D}.
\end{equation}
Thus, by taking the variation with respect to $\Psi^{D}$ to get
\begin{equation}\label{uniqIDw304}
\frac{D U^{D}}{DS} = \frac{m(S)_{D}}{m(S)}(G^{DC}- U^{D} U^{C}) + \frac{1}{2m(S)} R^{D}_{~CAB} S^{CAB}.
\end{equation}
And by taking the variation with respect to $\Psi^{\rho \delta}$ to obtain
\begin{equation}\label{uniqIDw305}
\frac{D\hat{S}^{C MN}}{Ds}=0.
\end{equation}
Furthermore, the spinning density tensor with a variable mass may be expressed in the following way:
\begin{equation}\label{uniqIDw306}
\frac{D \hat{S}^{CMN}}{DS} = ( \frac{m(S)_{W}}{m(S)}(G^{WC}- U^{W} U^{C}) + \frac{1}{2m(S)} R^{C}_{~ LAB} S^{LAB} ) S^{MN}.
\end{equation}
Accordingly, its corresponding Bazanski Lagrangian may be expressed as
\begin{equation}\label{uniqIDw307}
 L=  \hat{S}_{CMN} \frac{D \bar{\Psi}^{CMN}}{{D}{S}} + \bar{f}_{CMN}{\Psi}^{CMN} ,
\end{equation}
where $ \bar{f}^{CMN}= ( \frac{m(S)_{W}}{m(s)}(G^{WC}- U^{W} U^{C}) + \frac{1}{2m(S)} R^{C}_{~ } S^{WAB}) S^{MN}. $
By taking the variation with respect to $\Psi^{ABC}$ we obtain
\begin{equation}\label{uniqIDw308}
\frac{D \hat{S}^{ABC}}{Ds}= \bar{f}^{ABC} ,
\end{equation}
As we follow the same procedure as given in equation (5.6) we obtain the deviation spinning density equation to become
by taking the variation with respect to $\Psi^{ABC}$ we obtain
\begin{equation}\label{uniqIDw309}
\frac{D^{2} \Psi^{ABC}}{Ds^{2}}= (\bar{f}^{ABC})_{;D} \Psi^{D} + S^{Xi[BC}R^{A]}_{~~X MN}U^{M}\Psi^{N}.
\end{equation}
This equation may be applied in studying the stability of a variable spinning disk which may work to explain the effect of mass excess in a region orbiting a compact object. Such an illustration may give rise to examine the effect of dark matter in the accretion disk of a compact object.
\section{Spinning Density Tensor and Spinning Density  Deviation Tensor Equations: A Spinning Fluid }
In this section, we are going to suggest that the relation between a variable mass and a spinning fluid in the following way,
\begin{equation}\label{uniqIDw401}
\hat{S}^{CMN} = (p + \rho)(S)U^{C} S^{MN},
\end{equation}
where,
~~~~~~~$m(S) = (p + \rho)(S)$.\\ Accordingly, if the pressure is turning to be only parameter of $ S $, while the density is becoming constant. Owing to this suggestion the spin fluid behaves like a spinning variable mass, such that:
\begin{equation}\label{uniqIDw402}
\frac{d m(S)}{ dx^{p}} = \frac{d p (S)}{dX^{p}}.
\end{equation}
Such an equivalence may give rise to suggest the following Lagrangian,
\begin{equation}\label{uniqIDw403}
L= (p+\rho)(s)G_{MN}U^{\mu} \frac{D \Psi^{\nu}}{DS} + (p_{C} + \frac{1}{2} R_{CMND} S^{ND} U^{M} )\Psi^{C} + S_{AB} \frac{D \Psi^{AB}}{S}+P_{C}\Psi^{C}.
\end{equation}
Such that, by taking the variation with respect to $\Psi^{D}$ to get
\begin{equation}\label{uniqIDw404}
\frac{D U^{D}}{DS} = \frac{p_{C}}{(p+\rho)}(G^{DC}- U^{D} U^{C}) + \frac{1}{2(p+\rho)} R^{D}_{CAB} S^{CAB}.
\end{equation}
Also, by taking the variation with respect to $\Psi^{MN}$ to obtain
\begin{equation}\label{uniqIDw405}
\frac{D\hat{S}^{MN}}{DS}=0
\end{equation}
Consequently, the equation of a spinning fluid can be expressed in the following way:
\begin{equation}\label{uniqIDw406}
\frac{D\hat{S}^{CMN}}{Ds} = ( \frac{ p_{\sigma}}{(p+\rho)}(G^{W C}- U^{W} U^{C}) + \frac{1}{2(p+\rho)} R^{C}_{~WAB} S^{WAB}) S^{MN}.
\end{equation}
Accordingly, its corresponding Bazanski Lagrangian may be expressed as
\begin{equation}\label{uniqIDw407}
 L=  \hat{S}_{CMN} \frac{D \bar{\Psi}^{CMN}}{{D}{S}} + \bar{f}_{CMN}{\Psi}^{CMN} ,
\end{equation}
where,
 ~~~~~$ \bar{f}^{CMN}= ( \frac{p_{W}}{p+\rho}(G^{WC}- U^{W} U^{C}) + \frac{1}{2(p+\rho)} R^{C}_{~WAB} S^{WAB} ) S^{MN}. $
By taking the variation with respect to $\Psi^{ABC}$ we obtain
\begin{equation}\label{uniqIDw408}
\frac{D \hat{S}^{ABC}}{Ds}= \bar{f}^{ABC} ,
\end{equation}

By taking the variation with respect to $\Psi^{ABC}$ we obtain
\begin{equation}\label{uniqIDw409}
\frac{D^{2} \Psi^{ABC}}{DS^{2}}= (\bar{f}^{ABC})_{;D} \Psi^{D} + S^{X[MN}R^{C]}_{~~XMN}U^{A}\Psi^{B}.
\end{equation}
\subsection{Modified Forms of Spin Density in a Clifford Space}
  In this part, we suggest a modified form of spin density tensor in Riemannian geometry to be; \\
 (a)\underline{ $P^{A}= mU^{A}$}
\begin{equation}\label{uniqIDw410}
  S^{ABC} = \frac{1}{3!} (S^{BC} U^{A} + S^{CA} U^{B}+ S^{AB} U^{C}).
  \end{equation}

  Differentiating both sides using covariant derivative,
\begin{equation}\label{uniqIDw411}
  \frac{D{S^{ABC}}}{Ds} = \frac{1}{3!} (\frac{D S^{BC}}{Ds} U^{A} + S^{BC} \frac{D U^{A}}{Ds}+ \frac{S^{BA}}{Ds} U^{B}+S^{CA} \frac{D U^{B}}{Ds}+
  \frac{D S^{AB}}{Ds}U^{C}+ S^{AB} \frac{D U^{C}}{Ds} ).
  \end{equation}
  (i) For $ \frac{DU^{A}}{Ds} =0 $ and $ \frac{DS^{AB}}{Ds} =0$ and substituting in \eqref{uniqIDw411}, we obtain
\begin{equation}\label{uniqIDw412}
  \frac{DS^{ABC}}{Ds} =0.
  \end{equation}
  (ii) For $\frac{DU^{A}}{Ds}= \frac{1}{2m}R^{A}_{~BCD} S^{CD} U^{A} $  and $ \frac{DS^{AB}}{Ds} =0$, \eqref{uniqIDw411} can be rewritten as
\begin{equation}\label{uniqIDw413}
  \frac{DS^{ABC}}{Ds} =\frac{1}{3!}(\frac{1}{2m}R^{A}_{~EFG } S^{EG} U^{E} S^{BC}+ \frac{1}{2m}R^{B}_{~EFG} S^{FG} U^{E} S^{C A}+\frac{1}{2m}R^{C}_{~EFG} S^{FG} U^{E}S^{EF}),
  \end{equation}
  i.e.
 \begin{equation}\label{uniqIDw414}
  \frac{DS^{ABC}}{Ds} =\frac{1}{3!}( \frac{1}{2}(R^{A}_{~EFG } S^{BC}+ R^{B}_{~EFG} S^{BC}+R^{C}_{~EFG} S^{AB}) S^{EFG}),
  \end{equation}
  to become
\begin{equation}\label{uniqIDw415}
  \frac{DS^{ABC}}{Ds} =\frac{1}{2m}(R^{(A}_{~~EFG } S^{BC)}) S^{EFG},
  \end{equation}
  (b)\underline{$P^{A}\neq mU^{A}$}
\begin{equation}\label{uniqIDw416}
  S^{ABC} = \frac{1}{3!} (S^{BC} P^{A} + S^{BC} P^{B}+ S^{AB} P^{C}).
  \end{equation}
  Differentiating both sides using covariant derivative,
\begin{equation}\label{uniqIDw417}
  \frac{D{\bar{S}^{ABC}}}{DS} = \frac{1}{3!} (\frac{D S^{BC}}{Ds} P^{A} + S^{BC} \frac{D P^{A}}{Ds}+ \frac{DS^{CA}}{Ds} P^{B}+S^{CA} \frac{D P^{B}}{DS}+
  \frac{D S^{AB}}{Ds}P^{C}+ S^{AB} \frac{D P^{C}}{DS} ).
  \end{equation}
However, if $\frac{DP^{A}}{DS} =\frac{1}{2}R^{A}_{~BC E} S^{CE}U^{B}$ and $\frac{DS^{AB}}{Ds} = P^{A}U^{B}- P^{B}U^{A}$
  then, we obtain
  $$
  \frac{D{\bar{S}^{ABC}}}{DS} = \frac{1}{3!} ((P^{B} U^{C}- P^{C}U^{B}) P^{A} + \frac{1}{2}R^{A}_{~EFG} S^{EFG}S^{BC} + (P^{C}U^{A}-P^{A}U^{C})P^{B}
  $$
\begin{equation}\label{uniqIDw418}
  ~~~~~~~~~+ \frac{1}{2}R^{B}_{~EFG} S^{EFG}S^{CA}+(P^{A}U^{B}-P^{B}U^{A})P^{C}+  \frac{1}{2}R^{C}_{~EFG}S^{EFG}S^{AB} ),
  \end{equation}
  in which can be
\begin{equation}\label{uniqIDw419}
  \frac{D{\bar{S}^{ABC}}}{DS} = \frac{1}{12} S^{EFH}(S^{AB}R^{C}_{~EFH}+S^{BC}R^{A}_{~EFH}+S^{CA}R^{B}_{~EFH} ).
  \end{equation}
To get their corresponding deviation equation as similar as equation \eqref{uniqIDw409}.

\section{The Stability Problem of Spinning Density Using Spinning Deviation Poly-Vector}
It is well known that deviation vectors may work to express the stability of celestial objects. The role of employing the deviation vector is related to find out some scalrs independent of coordinate system. This technique has been propsed by Wanas and Bakry \cite{wanas}, and used in examining the stability of spinning object orbiting SgrA* by Kahil (2015)\cite{Kahil2015} and to be extended to spinning tensor deviation as shown by Kahil(2019)\cite{Kahil2019}. In this section, we are going to suggest the role of deviation poly-vector tensor as expressed in Clifford space to act similarly as in the case of the Riemannian geometry. This can be due to the following assumptions 
based on the deviation poly-vector.
$$
 \frac{D^2 }{D S^2}\Psi^ A=  R ^{A}_{.~BCD}~ U^{B}~ U^{C}~\Psi^{D} 
$$ to become
\begin{equation}\label{uniqIDw420}
\frac{D^2 }{D S^2}\Psi^ A = H^ {A}_{.~D}\Psi^{D} 
\end{equation}
i.e.
 $$ 
 H^ {A}_{.~D}:= R ^{A}_{.~BCD}~ U^{B}~ U^{C}
 $$
Also, for Spin deviation poly-tensor 
\begin{equation}\label{uniqIDw421}
\frac{D^2 }{D S^2}\Psi^ {AB} = S^{[~BE} R^{A]}_{.~EFD}\Psi^{D}U^{F} 
\end{equation},
$$
\frac{D^2 }{D S^2}\Psi^ {AB}= \tilde{H}^{AB}_{D}\Psi^{D},
$$
where,
$$
\tilde{H}^{AB}_{D}=S^{[~BE} R^{A]}_{.~EFD} U^{F}$$
Thus, in case of spin density  poly-tensor which is related to its corresponding stability tensor in the following way 
\begin{equation}\label{uniqIDw422}
\frac{D^2 }{D S^2}\Psi^ {ABC} = S^{[~BCE} R^{A]}_{.~EFD}\Psi^{D}U^{F} 
\end{equation},
to be expressed as
$$
\frac{D^2 }{D S^2}\Psi^ {ABC}= \tilde{H}^{ABC}_{D}\Psi^{D},
$$
where,
$$
\tilde{H}^{A}_{D}=S^{[~BE} R^{A]}_{.~EFD} U^{F}
$$
We figure out that the right hand side in equation (7.1) containing the deviation poly vector. This means that identifying the deviation of the poly vector 
will give rise to examine the stability of spinning tensors and spinning fluids as well as the generalized spinning density tensor. These equations are going to be examining stability of celestial objects wether are belonging to weak gravitational fields as in solar system or very strong graviataional fields as recognized in AGN that will be in our future work.
\section{Discussion and Concluding Remarks}
To sum up, the problem of motion in C-space becomes a paradigm shift towards identifying the behavior of objects in a deterministic way that can be detected  with the help of internal coordinates as well as their counterpart external coordinates. Such a tendency may inspire many scientist to reach to a unifed theory able to explain both micro-physics and macro-physics in a deterministic way.

Equations of motion for spinning objects as described in Clifford space give rise to extending to a spinning density object subject to different fields besides the gravitation field.
The use of imposing Clifford Space is to regard the internal coordinates that is associated with every poly-vector coordinate to express the effect of various fields than gravity that have a gauge property.

The problem of stability using the Spinning Deviation Tensors of Clifford space may gives rise its relationship with the deviation poly-vector of space-time.

 In the present work, we focused on spinning fluid tensor and its deviation one as a special case of applying spinning density tensor. In principle, a spinning fluid tensor is a special case of the spin density tensor with a skew-symmetry in the last two indices as expressed by Weyssenhoff tensor.  

 The spinning density tensor may play to express generalized case of spinning matter viable to express plasma physics nd heavy ion collision \cite{Gallegos2022}. In this perspective the 3rd rank skew symmetric poly-tensor turns to be specified to become $S^{ABC}$ a pure skew symmetric polytensor as defined in (6.10) an approach to relate the spin density tensor  a generalization of Weyssenhoff tensor .

Finally, it can be regarded that Clifford spaces may contain a specific version of Clifford Space of Gravity capable of embodying bi-gravity, and nonsymmetric gravitational field  theories for examining the behavior of matter in strong gravitational fields, that will be obtained in our future work!!   


\end{document}